\documentstyle[prb,array,multicol,aps]{revtex}
\newcommand{\keff}{K_2^{\hbox{\tiny eff}}}  
  
\begin{document}
\title{Absence of stable collinear configurations 
in Ni(001) ultrathin films: 
canted domain structure as ground state
}
\author{G. Gubbiotti,$^1$ G. Carlotti,$^{1,2}$ 
M. G. Pini,$^{3,*}$
P. Politi,$^{4,*}$ A. Rettori,$^{4,5}$ P. Vavassori,$^6$ M. Ciria,$^{7,8}$ 
and R. C. O'Handley$^8$}
\address{$^1$ Istituto Nazionale per la Fisica della Materia,
Unit\`a di Perugia, Via A. Pascoli, I-06123 Perugia, Italy} 
\address{$^2$ 
Dipartimento di Fisica dell'Universit\`a,Via Pascoli, I-06100 Perugia, Italy}
\address{$^3$ Istituto di Elettronica Quantistica, Consiglio Nazionale 
delle Ricerche, Via Panciatichi 56/30, I-50127 Firenze, Italy}
\address{$^4$ Istituto Nazionale per la Fisica della Materia,
Unit\`a di Firenze, Via G. Sansone 1, I-50019 Sesto Fiorentino (FI), Italy}
\address{$^5$ Dipartimento di Fisica dell'Universit\`a,
Via G. Sansone 1, I-50019 Sesto Fiorentino (FI), Italy}
\address{$^6$ 
Dipartimento di Fisica dell'Universit\`a,
Istituto Nazionale per la Fisica della Materia,
Via Paradiso 12, I-44100 Ferrara, Italy}
\address{$^7$ Departamento de Fisica de la Materia 
Condensada, Universidad de Zaragoza, C/Pedro Cerbuna
13, 50009 Zaragoza, Spain}
\address{$^8$
Department of Materials Science and Engineering, 
Massachusetts Institute of 
Technology Cambridge, MA 02139, USA}
\date{\today}
\maketitle
\begin{abstract}
Brillouin light scattering (BLS) measurements
were performed for (17-120)\AA-thick Cu/Ni/Cu/Si(001) films.
A monotonic dependence of the frequency of the uniform mode
on an in-plane magnetic field $H$ was observed 
both on increasing and on decreasing $H$ in the range (2-14) kOe,
suggesting the absence of a metastable collinear ground state.
Further investigation by magneto-optical vector
magnetometry (MOKE-VM) in an unconventional 
canted-field geometry provided evidence for a
domain structure where the magnetization is canted with respect
to the perpendicular to the film.
Spin wave calculations confirm the absence of stable collinear
configurations.
\end{abstract}
\begin{multicols}{2}
\section{introduction}
The determination of the magnetic phase of thin films can require 
a subtle analysis owing to the presence of many different 
competing anisotropies. Furthermore, metastable phenomena
can be present, so that the magnetic history of the sample
must be considered. For example, when the magnetization is
perpendicular to the film plane, a monodomain state is suggested
from remanence of polar Kerr loops,\cite{Qiu} 
even though a striped domain configuration is the 
ground state.\cite{Yafet}
However, it is worthwhile to note that, in general, 
all the investigated films present either perpendicular
($\theta=0$) or in-plane ($\theta=\pi/2$) magnetization ($\theta$ 
is the angle between the magnetization and $z$, the normal to the 
film plane); a phase with canted ($0<\theta<\pi/2$) magnetization 
has been usually observed close to the reorientation transition
between perpendicular and in-plane magnetization in a very narrow range
of film thickness, $t$, and/or temperature,
$T$.\cite{Ha,Ibach,Bischof,Allenspach,Farle}
This phase is usually related to a domain structure as 
experimentally observed by microscopic 
techniques.\cite{Ibach,Bischof,Allenspach}
Canting of the magnetization has been observed in a class of
rather different systems as well: multi\-layered structures of Fe/Tb, 
for example.\cite{FeTb} 

It is generally agreed that epitaxial Cu/Ni/Cu/Si(001) films 
present a collinear phase magnetized perpendicularly 
to the film plane in a wide range of Ni thickness, 
17 \AA$\le t \le$120 \AA.\cite{Ha,Farle,Bochi,IEEE}
Assuming the perpendicular state to be stable, the frequency $\Omega_0$
of the uniform mode versus $H$, an external magnetic field 
applied within the film plane, is predicted 
to have a minimum at a critical field $H_c$.\cite{Dutcher}
In a previous paper,\cite{JAPnoi} 
we reported Brillouin light scattering (BLS)
experiments in Cu/Ni/Cu/Si(001) films which, in the
thickness range $17$ \AA $\le t \le80$ \AA, showed a {\it monotonic}
decrease of the frequency of the uniform mode on decreasing the
in-plane field down to 2 kOe, with a clear change in slope 
at field values $H^* \approx (5$-$8)$ kOe, depending on 
the film thickness. 
(A similar behavior was observed also in other systems:
Co/Au ultrathin films,\cite{Falco} 
Co/Pd multilayers,\cite{Harzer}
and Co/Au multilayers.\cite{Albertini})
We attributed such
a discrepancy between our data and theoretical predictions\cite{Dutcher}
to the breakdown of the collinear ground state state and to the onset of a 
domain configuration.\cite{JAPnoi} As a matter of fact, spin-wave 
theory\cite{Bruno,Erickson,JMMMnoi} predicts such an instability
at a field value $H_c^> = H_c+\Delta H_c$ 
(see Section V for the definitions of $H_c$ and $\Delta H_c$)
and at a finite wavevector $k_{\Vert}$, 
whose inverse sets the initial size of the domains.
In our previous paper\cite{JAPnoi} 
strong evidence for the presence of domains 
was provided by magneto-optical vector-magnetometry (MOKE-VM) data.
A direct visualization of up and down perpendicular domains in these specimens
for zero applied field was presented by Hug {\it et al.};\cite{Hug}
unfortunately, our present attempts to visualize -  using
magnetic force microscopy - the domain structure
while applying an in-plane magnetic field were unsuccessful due
to the very weak magnetic signal from the sample and to 
the interference with the strong external field. 

If the system is saturated in the $z$ (perpendicular) direction and
the in-plane field $H$ is increased from 0 to $H_c$, 
a collinear ground state
with a canting angle $\theta$ increasing from $0$ to $\pi/2$
is expected, and the spin-wave
frequency is predicted to decrease until, at a field value
$H_c^<\approx H_c-\Delta H_c/2$, a ``lower" instability 
(quite similar to the ``upper" one at $H_c^>$) 
is expected.\cite{Erickson} The fact that, on decreasing $H$,
the Brillouin light scattering data~\cite{JAPnoi} did {\it not} 
show any indication for the low field phase 
might be interpreted as due to a metastability 
phenomenon.

The aim of the present paper is to definitively investigate the possible
occurrence of the above mentioned metastability and to achieve a detailed
understanding of the ground state and of both the static and dynamic 
magnetization properties.
New BLS data taken by increasing the in-plane field $H$ 
are presented and discussed. 
Moreover, MOKE-VM data in a new canted-field
geometry are presented in order to analyze the collinear ground state.

\section{Brillouin light scattering data}

Our new measurements of BLS spectra are reported 
in Fig. 1 for the Cu/Ni/Cu/Si(001) sample with
Ni thickness $t=60$ \AA$~$ for several values of the 
external magnetic field $H$ applied within the film plane
(similar results are obtained for other
thicknesses in the range 17\AA-80\AA). 
The details of the film growth and structural characterization 
have been published elsewhere,\cite{growth} but a brief description 
is necessary here. The films were grown in a molecular beam epitaxy (MBE) 
chamber by e-beam evaporation. The base pressure was less than 
$2.0$ $10^{-10}$ Torr 
and rose in the low $10^{-8}$ Torr range during deposition
of both Nickel and Copper layers. 
The growth rates for the Copper and Nickel layers were 1.0$~$\AA/s.
The copper buffer layers for all the samples were annealed 
{\it in situ} to about 150$^{\circ}$C for eight minutes
to improve the flatness of the buffer layer surface.
Unlike the case of Cu on Si(001), the interface betwen Ni(001) 
and Cu(001) remains sharp at room temperature. 
It has been shown by  Chen et al. \cite{chen} 
that the Ni on Cu is thermally stable against 
interdiffusion up to 200$^{\circ}$C.
Reflection high energy electron diffraction (RHEED) and synchrotron 
x-ray diffraction show high quality epitaxial Nickel films.
BLS measurements were carried out in the 
backscattering geometry at the GHOST 
laboratory in Perugia\cite{www} 
using a Sandercock-type tandem Fabry-Perot 
interferometer in a  (3+3)-pass configuration.\cite{Sand}
About 150 mW of P-polarized light, from an Ar$^+$-ion 
laser operated in single longitudinal 
mode on the 5145-\AA~line, was focused onto the sample surface, at an 
incidence angle of 45$^{\circ}$, by a camera objective of numerical aperture 
2 and focal length 50 mm. The sample was placed between the poles 
of an electromagnet used to produce a d.c. magnetic field, with a 
maximum intensity of 15.0 kOe, applied parallel to the film surface 
and perpendicular to the plane of incidence of light. Since light 
scattered by magnons has its 
plane of polarization rotated through 90$^{\circ}$, 
an analyzer was used to remove unwanted back-reflections from the 
objective lens and light scattered by acoustic phonons. The spectra, 
recorded at room temperature, were stored in 512 channels, each with 
a gate length of 1 ms. A few thousand interferometer scans were necessary 
to record spectra with a good signal-to-noise ratio.

\section{Model}

From the measured spectra in Fig. 1, one obtains the field
dependence of the frequency of the Damon-Eschbach mode 
reported in Fig. 2.
A detailed theoretical analysis of these data 
is performed in terms of the following spin Hamiltonian
\begin{eqnarray}
{\cal H}= &-&J\sum_{i,\delta} {\bf S}_i \cdot {\bf S}_{i+\delta}
- g \mu_B {\bf H} \cdot \sum_i {\bf S}_i 
- K_2 \sum_i (S_i^z)^2
\cr
&+& {1\over 2} K_{4}^{\perp} \sum_i (S_i^z)^4
+ {1\over 2} K_{4}^{\Vert} \sum_i [(S_i^x)^4+(S_i^y)^4]
\cr
&+& {1\over 2} (g \mu_B)^2 \sum_{i\ne j} 
\left\lbrack 
{ { {\bf S}_i \cdot {\bf S}_j }\over { {\bf r}_{ij}^3 } }
-3 { { ({\bf S}_i \cdot {\bf r}_{ij})({\bf S}_j \cdot {\bf r}_{ij}) }\over 
{ {\bf r}_{ij}^5} }
\right\rbrack
\end{eqnarray}
\noindent where $J>0$ is the nearest neighbour ferromagnetic exchange 
constant; $H$ is an external field;
$K_2>0$ is a uniaxial anisotropy favouring the $z$ direction, normal
to the film; $K_{4}^{\perp}>0$ and $K_{4}^{\Vert}$
are quartic anisotropies:\cite{Ha,notaFarle} the former favours
the film plane, $xy$; the latter, when 
the spins are in plane, favours them to lie along the in-plane
diagonal [110]; finally, the last term in Eq. (1)
is the magnetostatic dipole-dipole interaction.
The presence of a quartic in-plane anisotropy is required since
the BLS data, performed for field applied in plane 
along the easy [110] and the hard [100] direction respectively,
show a small difference in frequency (see the inset in Fig. 2). 
For sufficiently high fields ($H> H_c^>$),
the spins are aligned in plane along the field direction and  
the frequency of the uniform mode is\cite{Farle}
\begin{mathletters}
\begin{equation}
\label{easy}
\Omega_0\vert_{[110]}=\gamma
\sqrt{(H-H_{\keff}-{1\over 2}H_{K_{4}^{\Vert}})
(H+H_{K_4^{\Vert}})}
\end{equation}
\begin{equation}
\label{hard}
\Omega_0\vert_{[100]}=\gamma
\sqrt{(H-H_{\keff}-H_{K_{4}^{\Vert}})
(H-H_{K_{4}^{\Vert}})}
\end{equation}
\end{mathletters}
for $H$ along the easy and hard in-plane axis, 
respectively ($\gamma$ is the gyromagnetic factor). 
Here above and in the following we put
$H_{\keff}=H_{K_2}-H_{dip}$, where
$g \mu_B H_{K_2}=2 K_2 S$ is the uniaxial anisotropy field 
and $H_{dip}=4 \pi M_0 c_1$ the dipolar field 
($M_0$ is the bulk magnetization and 
$c_1=f/\sqrt{2}$ is a thickness dependent 
coefficient\cite{review,notaYE});
$g\mu_B H_{K_{4}^{\Vert}}=2K_{4}^{\Vert}S^3$ and
$g\mu_B H_{K_{4}^{\perp}}=2K_{4}^{\perp}S^3$ are the 
quartic anisotropy fields. The former is estimated 
to be $H_{K_{4}^{\Vert}} \approx 1$ kOe from the
frequency difference in the BLS data
($\Delta \nu \approx (2$-$3)$ GHz for $H \gtrsim 10$ kOe),
while a fit of $\Omega_0(H)$ at high fields along [110],
using Eq. (\ref{easy}), yields
an estimate for the effective anisotropy field 
$H_{\keff}\approx 2.5$ kOe. Finally, we define the
exchange field $g \mu _B H_{ex}=2JS$ which - though not 
appearing in Eqs. (\ref{easy},\ref{hard}), since they  
are written for zero wavevector -
will be useful in the following.

The results of our new BLS measurements in Fig. 2 
clearly show that almost the {\it same} curve 
$\Omega_0$ versus $H$ is obtained on increasing
as well as on decreasing the in-plane field $H$ 
in the range (2-14) kOe.\cite{nota2kOe}
We conclude that {\it 
the collinear perpendicular state is a very weakly metastable 
state or that it is plainly unstable}, {\it i.e.} the system does not 
present a perpendicularly magnetized phase. 
Even more important, BLS data exclude a ground state with domains
magnetized perpendicularly to the film plane as well.
In fact, for $H=0$ and negligible quartic anisotropies,
the energy gap of the spin-wave excitations with respect to
a ground state with striped domains perpendicularly magnetized
is estimated\cite{Ramesh} to be 
\begin{equation}
\Omega_0\vert_{\perp,stripe}=
\gamma (H_{K_2}-H_{dip}N_{zz})
\end{equation}
\noindent where $N_{zz}<1$ is the 
demagnetization factor.\cite{notaNzz}
Thus, in a perpendicular striped domain structure, the gap 
would be {\it higher} than in the case of spin waves excited from
a uniform perpendicular ground state ({\it i.e.} with $N_{zz}=1$)
\begin{equation}
\Omega_0\vert_{uniform}=\gamma(H_{K_2}-H_{dip}).
\end{equation}
\noindent Since, using the value of $H_{\keff}$ deduced before, 
the latter frequency is estimated to be at least 7 GHz, 
the hypothesis
of a perpendicular striped domain structure in the low field region
($H<2$ kOe) appears to be inconsistent with the high field 
($H>10$ kOe) results.
This problem can be solved assuming the system to have a quartic 
perpendicular anisotropy $K_{4}^{\perp}>0$,
{\it i.e.} competing with $\keff$ and leading to a canted 
collinear configuration for $H=0$. Neglecting the small in-plane
quartic anisotropy, the zero-field canting angle $\theta_0$ is given by 
$\cos^2 \theta_0=H_{\keff}/H_{K_{4}^{\perp}}$ (see Eq. (\ref{angle}) 
in the Appendix).
In this way, the problem 
of the inconsistency between low and high field BLS data is
removed because, for $H \to 0$, the spectrum of the excitations 
with respect to such a canted collinear ground state has
a Goldstone mode ($\Omega_0=0$) as a consequence of the 
rotational invariance around the $z$ axis.
In the presence of a small quartic in-plane anisotropy, 
the gap at zero field is slightly greater than zero, but 
small enough to make our argument still valid.

\section{MOKE-VM data}

In order to reveal the possible occurrence 
of a canted magnetization,\cite{canting} we use the magneto-optic 
Kerr vector-magnetometry (MOKE-VM) technique\cite{Vavassori} 
in a novel canted-field geometry.
The field is applied within the plane formed by $z=[001]$, 
the normal to the film, and $x=[110]$, the in-plane easy axis;
the field direction formed a canting angle $\theta_H$
with $z$.
In Fig. 3a,b we report the MOKE-VM data for the 
polar ($M_z$) and the longitudinal ($M_x$)
component of the magnetization, respectively.
In the present experimental configuration we did not detect
any transversal component ($M_y$) in the 
whole investigated field range. 
In Fig. 3c, we show the field dependence of the calculated
magnetization modulus $M=\sqrt{M_x^2+M_z^2}$, 
normalized to the saturation value $M_{sat}$.

The different curves refer to different
orientations of the applied magnetic field, with 
$\theta_H$ ranging between 0$^{\circ}$ and 20$^{\circ}$. 
The perpendicular component $M_z$ shows a remanence
in zero field, the greater the lower is $\theta_H$.
Also the parallel component $M_x$ is found to be nonzero 
in zero applied field; moreover, when the applied 
magnetic field is sufficiently high ($|H|>4$ kOe),
$M_x$ is found to decrease with increasing $H$
for $\theta_H \le 11^{\circ}$ and to increase 
for $\theta_H=20^{\circ}$.
Therefore, these observations call for a canted ground state 
configuration in zero field with a canting angle 
$\theta_0$ comprised between 11$^{\circ}$ and 20$^{\circ}$.
For the canted-field geometry of the MOKE-VM
experiment one has (see Eq. (\ref{angle}) in the Appendix)
\begin{equation}
\label{theta0}
\cos^2 \theta_0=
(H_{\keff}+{1\over 2}H_{K_{4}^{\Vert}})/
(H_{K_{4}^{\perp}}+{1\over 2}H_{K_{4}^{\Vert}}).
\end{equation}
\noindent Hence, putting $\theta_0\approx 15^{\circ}$,  
we obtain $H_{K_{4}^{\perp}} \approx  2.7$ kOe.
Concerning the magnetization modulus, we observe that it is 
always decreasing as the field is decreased, mainly owing 
to the the decrease of the $M_z$ component. This is 
a strong evidence for
a slightly canted ground state with a 
domain structure for the perpendicular component
of the magnetization: ${\bf M}=(M_x,0,\pm M_z)$.

\section{Discussion}
Now, after having estimated  all the 
Hamiltonian parameters which determine the frequency
of the uniform mode ({\it i.e.} $H_{\keff}\approx 2.5$ kOe, 
$H_{K_4^{\Vert}}\approx 1$ kOe and
$H_{K_4^{\perp}}\approx 2.7$ kOe) 
we note that, using these values,
the BLS data are well reproduced (see Fig. 2) 
only in a very high field region ($H \gtrsim 10$ kOe).
In order to discuss the possible arising of
instabilities,\cite{Bruno,Erickson,JMMMnoi}
we need to generalize the calculation 
of the spin-wave frequency gap
to a finite wavevector, 
$\Omega_0({\bf k}_{\Vert})$.
For doing that, we specialize to the in-plane easy axis
direction [110].

For long wavelength, the spin-wave acoustic mode
(which is identified with the Damon-Eschbach 
mode in the Brillouin spectrum) 
of an ultrathin film with $N$ planes takes the approximate form
\begin{equation}
\label{approxi}
\Omega_0({\bf k}_{\Vert})\approx \gamma \sqrt{
A_1({\bf k}_{\Vert})~A_2({\bf k}_{\Vert}) } 
\end{equation}
where the full expressions 
of $A_1({\bf k}_{\Vert})$ and $A_2({\bf k}_{\Vert})$ 
are given in the Appendix.
This expression is approximated in the sense that it 
is the generalization of the spin-wave dispersion relation 
of the monolayer\cite{YKG,Bruno,Erickson} 
to the acoustic mode of an ultrathin film with $N$ planes.\cite{JMMMnoi}
Since $A_1({\bf k}_{\Vert})$ is always strictly positive,
the only reason for a field-induced instability at finite wavevector 
can be the vanishing of $A_2({\bf k}_{\Vert})=\alpha_0-\alpha_1 (k_{\Vert}a)
+\alpha_2 (k_{\Vert}a)^2$ (where $a$ is the square lattice constant 
and the $\alpha_i$'s are known functions of the physical parameters, 
see Eq. (\ref{A2}) in the Appendix).
The minimum of $A_2({\bf k_{\Vert}})$ is obtained for
$k_{\Vert}a=k_{\Vert}^{(m)}a=\alpha_1/(2\alpha_2)$
and we obtain 
$A_2(k_{\Vert}^{(m)})=\alpha_0-\alpha_1^2/(4\alpha_2)$: 
if this quantity is negative, 
the spin wave frequency becomes pure imaginary, signaling the
instability of the uniformly magnetized canted state. 
The instability appears when $A_2(k_{\Vert}^{(m)})=0$: this
condition defines the values of the 
upper ($H_c^>$) and lower ($H_c^<$) critical fields. 
In the high field regime we find
$H_c^>=H_c+\Delta H_c$, with 
$H_c=H_{\keff}+{1\over 2} H_{K_4^{\Vert}}$ and
$\Delta H_c=(H_{dip}N)^2/(16 f^2 H_{ex})$,
and $k_{\Vert}^{(m)}a=(H_{dip}N)/(4 f H_{ex})$:
the uniform ground state breaks into domains of size 
$\approx 1/k_{\Vert}^{(m)}$. 

The BLS data show the existence of two relevant fields:
at $H\approx 9$ kOe experimental data 
start deviating from the theoretical spin wave frequency
and at $H^* \approx 6$ kOe there is a clear change in their slope.
The higher field is interpreted to signal
the arising of a domain structure, which
makes the experimental $\Omega_0$ deviate from the value predicted for 
a collinear in-plane phase: it is therefore identified as the
field $H_c^>$. 
Instead, the changement in the slope at a field
$H^*\approx 6$ kOe should signal that a well defined domain
structure is now established, whose size may strongly
interfere with the light used in the BLS experiment.

We can sum up our interpretation as follows.
At high fields we have an in-plane collinear phase and we start
decreasing $H$. At $H=H_c^>\approx 9$ kOe,
the collinear phase is
unstable against the appearance of a domain structure:
spins acquire a very small $z$ component that changes sign from
a domain to a neighbouring domain.
This instability corresponds to the vanishing of
$A_2({\bf k}_{\Vert})$ for a value $k_{\Vert}^{(m)}$ which is just the
inverse of the domain size.
With further decreasing $H$, the domain structure becomes more and
more defined and the domain size {\it increases}.

On the other side, the ``lower"
instability occurs for $H_c^< \approx H_{c}-\Delta H_c/2$.
In our present interpretation, the quantity $\Delta H_c$
is taken as a free parameter\cite{free}
and estimated, from the high field results, 
to be $\approx 6$ kOe. Consequently, 
since $H_c \approx 3$ kOe,
the ``lower" instability is expected to occur already for 
$H \approx 0$. We observe that such a conclusion about
the instability of a collinear canted ground state 
is strongly supported by our MOKE-VM results in canted field geometry: 
even a relatively high field $H\approx 4$ kOe applied 
nearly along the easy 
magnetization axis ($\theta_H=11^{\circ}$ or $20^{\circ}$) 
is unable to give the saturation value
of the magnetization modulus, see Fig. 3c.

In conclusion, new BLS measurements, performed upon increasing 
the in-plane field $H$, allowed to definitely exclude
a metastable collinear state for epitaxial Cu/Ni/Cu/Si(001) films 
with thickness in the range (17-80)\AA.$~$
This feature is confirmed by spin-wave calculations.
The MOKE-VM data in canted-field geometry suggested the
presence of a domain structure where the magnetization is 
slightly canted ($\theta \approx 15^{\circ}$) 
with respect to the film normal. 
We are confident that our results can be useful to explain
spin wave anomalies previously observed in other systems, 
such as Co/Au ultrathin films\cite{Falco}, as well as
Co/Pd multilayers\cite{Harzer} and Co/Au 
multilayers.\cite{Albertini,notaAlb}

\appendix
\section{Spin-wave dispersion}

The spin-wave dispersion relation of the acoustic mode frequency
of an fcc $N$-planes ultrathin film, described by Hamiltonian (1),
is found to be, in the low wavevector limit
\begin{equation}
\label{general}
\Omega_0({\bf k}_{\Vert})\approx\gamma 
\sqrt{
A_1({\bf k}_{\Vert})~A_2({\bf k}_{\Vert})-[{\rm Im}B({\bf k}_{\Vert})]^2
}
\end{equation}
where
\begin{mathletters}
\begin{eqnarray}
A_1({\bf k}_{\Vert})&=&\Big\{
H [\sin \theta \sin\theta_H \cos(\varphi-\varphi_H) 
+\cos\theta\cos\theta_H]
\cr
&+& H_{\keff} \cos^2 \theta
- H_{K_{4}^{\perp}} \cos^4 \theta
\cr
&+&{1\over 4}H_{K_{4}^{\Vert}} \sin^2 \theta 
[3\cos^2 \theta -\cos 4 \varphi(3+\sin^2 \theta)]\Big\}
\cr
&+&\Big\{{{N}\over {2f}} H_{dip}
\sin^2(\varphi-\varphi_k)
\Big\}\cdot(k_{\Vert}a) 
\cr
&+&H_{ex}\cdot(k_{\Vert}a)^2 
\end{eqnarray}
\begin{eqnarray}
\label{A2}
A_2({\bf k}_{\Vert})&=& 
\alpha_0 - \alpha_1 \cdot(k_{\Vert} a)+\alpha_2 \cdot(k_{\Vert}a)^2
\cr
&=&
\Big\{ H[ \sin \theta \sin \theta_H \cos(\varphi-\varphi_H) 
+\cos\theta \cos\theta_H]
\cr
&+& H_{\keff} \cos(2 \theta)
- H_{K_{4}^{\perp}} \cos^2 \theta (1- 4\sin^2 \theta)
\cr
&-&{1\over 4}H_{K_{4}^{\Vert}} \sin^2 \theta (1-4\cos^2\theta)
(3 + \cos 4 \varphi) \Big\}
\cr
&-&\Big\{
{{N}\over {2f}} H_{dip}[\sin^2\theta-\cos^2\theta
\cos^2(\varphi-\varphi_k)] \Big\}
\cr
&\cdot& (k_{\Vert}a)
\cr
&+&H_{ex}\cdot(k_{\Vert}a)^2 
\end{eqnarray}
\begin{eqnarray}
{\rm Im}B({\bf k}_{\Vert})&=&\Big\{
-{3\over 4} H_{K_{4}^{\Vert}} \sin^2 \theta \cos\theta \sin 4 \varphi
\Big\}
\cr
&-&  \Big\{{{N}\over {4f}} H_{dip} \cos\theta
\sin[2(\varphi-\varphi_k)]
\Big\}\cdot (k_{\Vert}a)
\end{eqnarray}
\end{mathletters}

\noindent The expression for $\Omega_0({\bf k}_{\Vert})$ 
is approximated in the sense that it 
is the generalization of the spin-wave dispersion relation 
of the monolayer\cite{YKG,Bruno,Erickson} 
to the case of an ultrathin film with $N$ planes:\cite{JMMMnoi}
for the latter case, it represents the energy of the {\it acoustic}
mode of the film, which is measured in a BLS experiment. 
We notice that the 
effect of the finite number of planes manifests itself as a
prefactor $N$ in the terms containing $H_{dip}$ and linear 
in the wavevector $k_{\Vert}$.
In the previous equations, ${\bf k}_{\Vert}$
is the in-plane two-dimensional wavevector 
forming an angle $\varphi_k$ with the in-plane 
$x=[100]$ crystallographic axis; 
$a=a_0/\sqrt{2}$ is the constant of the square lattice 
on the (001) surface expressed in terms of 
$a_0$, the constant of the bulk fcc lattice;
$\theta$ and $\varphi$ denote the 
polar coordinates of the magnetization taking $z$
(the normal to the film plane) as polar axis, while
$\theta_H$ and $\varphi_H$ are
the polar coordinates of the applied magnetic field.
Dipolar sums were evaluated in the limit of small 
${\bf k}_{\Vert}$ using an Ewald-type summation 
method.\cite{YKG,Bruno,Erickson,JMMMnoi}
The dipolar field is $H_{dip}=4\pi M_0 c_1$,
where $M_0=4 g \mu_B S/a_0^3$ is the magnetization 
of the bulk fcc lattice and $c_1=f/\sqrt{2}$ is 
a thickness dependent coefficient.\cite{review,notaYE}
The other fields are defined in the text 
after Eqs. (\ref{easy},\ref{hard}).

In the general case, the ground state configuration 
$(\theta,\varphi)$ is obtained solving the system 
\begin{mathletters}
\begin{eqnarray}
\label{theta}
&[&H_{K_4^{\perp}}+{1\over 4}H_{K_4^{\Vert}}(3+\cos 4\varphi)]
\sin^3 \theta \cos\theta \cr
-&[&H_{K_4^{\perp}}-H_{\keff}]\sin \theta \cos \theta
\cr
-H&[&
\cos\theta\sin\theta_H \cos(\varphi-\varphi_H)
-\sin\theta \cos\theta_H
]=0
\end{eqnarray}
\begin{equation}
\label{phi}
H\sin\theta_H \sin(\varphi-\varphi_H)-{1\over 2}
H_{K_4^{\Vert}} \sin^3 \theta \sin 4 \varphi=0
\end{equation}
\end{mathletters}

\noindent When the external 
magnetic field is applied in plane 
along the easy direction
($\theta_H=\pi/2$, $\varphi_H=\pi/4$),
Eq. (\ref{phi}) is satisfied by $\varphi=\pi/4$,
while Eq. (\ref{theta}) becomes 
\begin{eqnarray}
\label{angle}
\cos\theta 
[(&H&_{K_4^{\perp}}
+{1\over 2}H_{K_4^{\Vert}})\sin^3 \theta \cr
-(&H&_{K_4^{\perp}}-H_{\keff})\sin\theta -H]=0
\end{eqnarray}
Hence, the zero-field canting angle reported in Eq. (\ref{theta0}) 
is obtained, provided that $H_{K_4^{\perp}}>H_{\keff}$.
Moreover we observe that, for $\varphi=\varphi_H=\pi/4$, 
ImB$({\bf k}_{\Vert})$ reduces to a term linear in $H_{dip}$
and in the wavevector: thus its square can safely be neglected 
in Eq. (\ref{general}), so that Eq. (\ref{approxi}) 
is obtained.
Finally, it is worth noticing that 
in the case of zero in-plane anisotropy,
$H_{K_4^{\Vert}}=0$, one recovers for 
$A_1({\bf k}_{\Vert})$ and $A_2({\bf k}_{\Vert})$ the expressions 
previously found by Erickson and Mills.\cite{Erickson}

\begin{figure}
\caption{
Brillouin light scattering spectra for several values of
the external field $H$, applied in-plane along the [100] 
direction. The Ni film was 60 \AA$~$thick, and data were 
obtained using the backscattering configuration; see text
for the experimental details.
}
\end{figure}

\begin{figure}
\caption{
Brillouin light scattering data for the spin wave frequency 
$\Omega_0$
of a $t=60$ \AA-thick Ni film versus the intensity 
of a magnetic field $H$ applied along the in-plane easy axis [110],
on increasing $H$ (solid squares) as well as
on decreasing it (solid circles). The solid line is the 
spin wave frequency, calculated from Eq.(\protect\ref{easy})
assuming a uniform in-plane magnetization. In the inset 
we compare the $\Omega_0(H)$ data measured
for $H$ applied along the easy axis [110] (solid symbols) 
and for $H$ along the hard axis [100] (open symbols).
}
\end{figure}
\begin{figure}
\caption{
Magneto-optic vector magnetometry data in a canted field 
geometry, showing the field dependence of: a) the polar 
($M_z$) component of the magnetization, 
b) the longitudinal ($M_x$) component,
c) the magnetization modulus $M=\sqrt{M_x^2+M_z^2}$.
All quantities are normalized to the saturation value
$M_{\rm sat}$.
The field was applied at different angles ($\theta_H=0^{\circ},5^{\circ},
11^{\circ},20^{\circ}$)
with the film normal, $z$.
}
\end{figure}

\end{multicols}

\begin{references}
\bibitem{notaFarle}
We observe that, in our notations,
the signs of the quartic anisotropies $K_{4}^{\perp}$ 
and $K_{4}^{\Vert}$ are reversed
with respect to those in Farle's paper.\cite{Farle}
\bibitem{notaYE}
For a monolayer, the coefficient $c_1$ was estimated\cite{Yafet,Erickson}
to be 0.762, while for specimens as thick as the investigated ones, its 
value is very nearly equal to 1.
\bibitem{nota2kOe}
The field dependence below 2 kOe could not be studied because
the central line overwhelmes the inelastic spin-wave mode.
\bibitem{notaNzz}
$N_{zz}$ depends on 
the aspect ratio $r=L/t$ ($L$ is the width of the stripes,
$t$ is the film thickness) and the ratio between the 
anisotropy field $H_{K_2}$ and the dipolar field $H_{dip}$:
in the limit of a uniformly magnetized sample ($r \to \infty$) 
one has $N_{zz} \to 1$.
\bibitem{canting}
The canting of the magnetization vector in systems like Fe/Tb
(see Ref.\protect\cite{FeTb}) has been observed using
a M\"ossbauer technique, which does not require an applied field to probe
magnetization.
\bibitem{free}
This can be justified since, in such ultrathin films,
the Hamiltonian parameters are not exactly known 
and even the film thickness is a nominal parameter.
\bibitem{notaAlb}
It is worthwhile noticing that in Co/Cu multilayers
a direct visualization of the domain structure using
magnetic force microscopy was possible.\cite{Albertini}
\end{references}
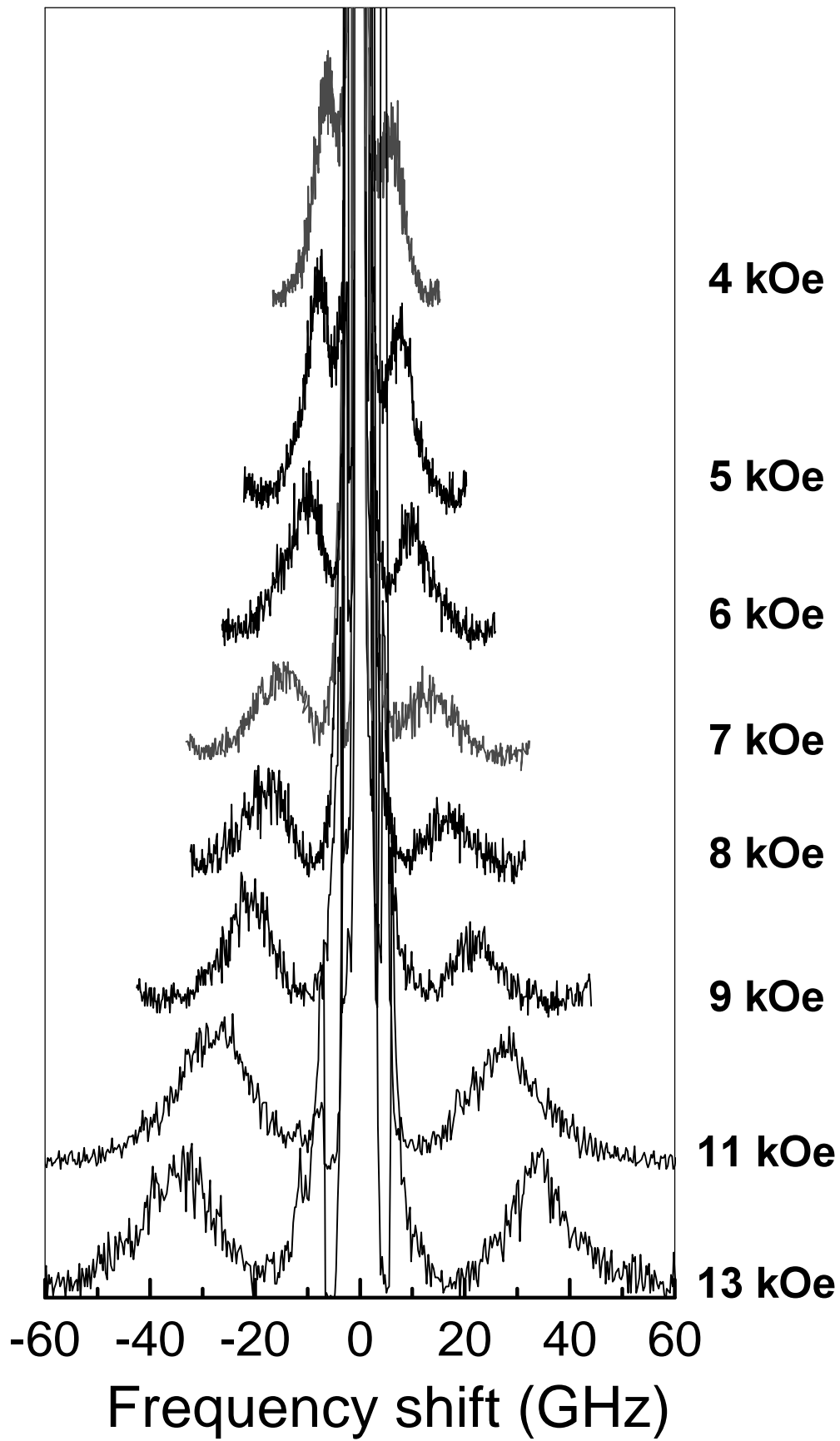
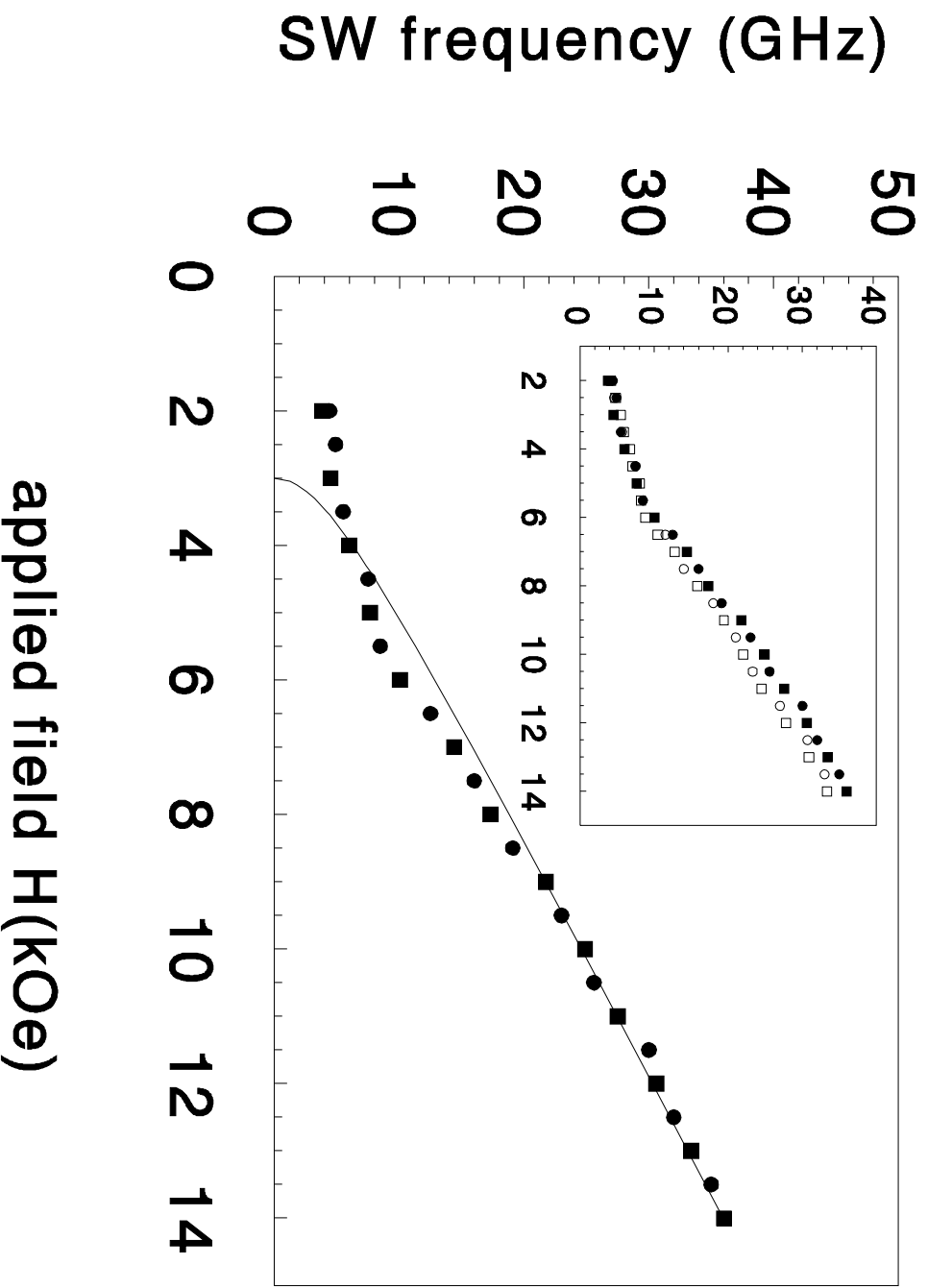
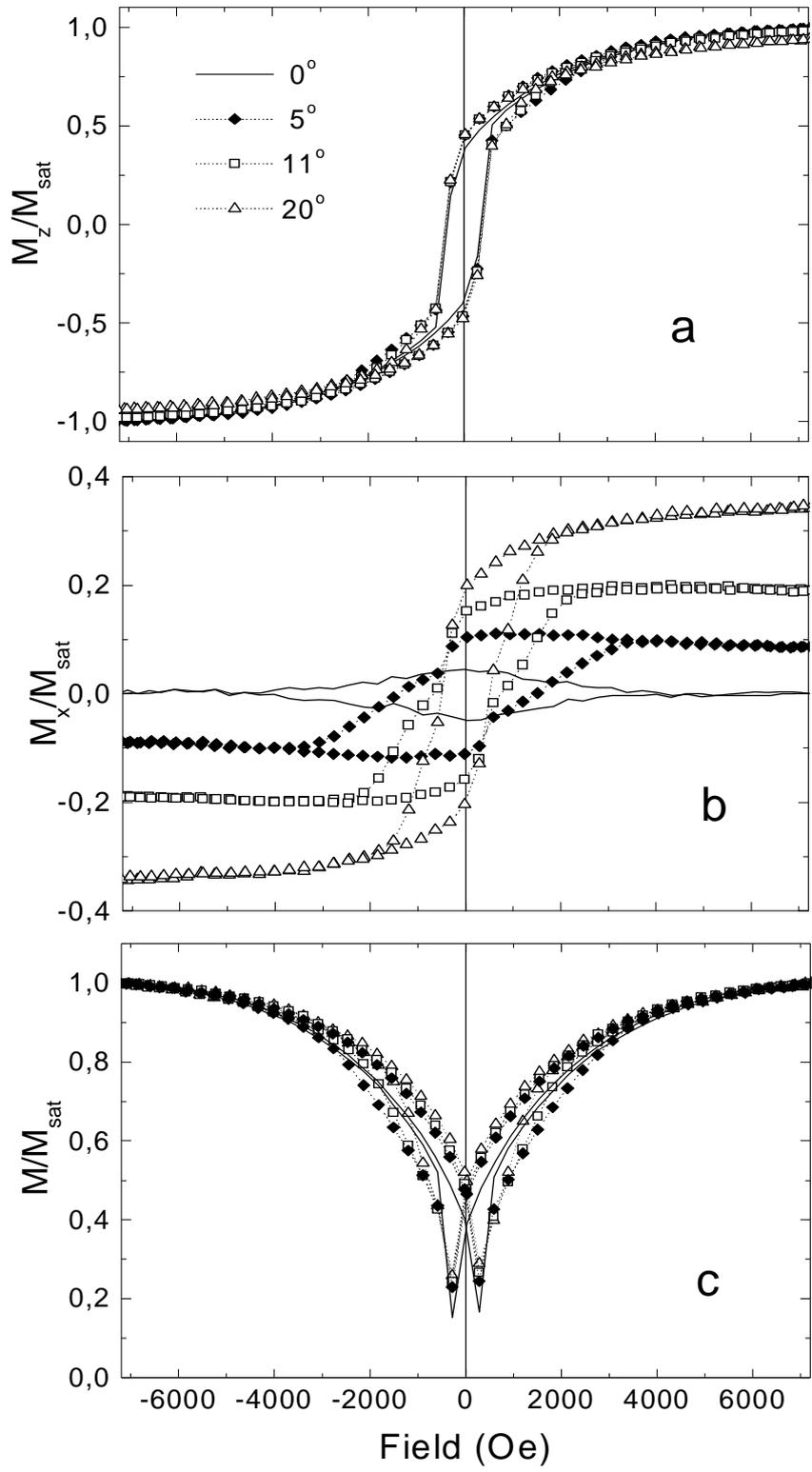
\end{document}